\documentclass[prd,aps,10pt,showkeys,nofootinbib,twocolumn]{revtex4-1}

\usepackage{amsmath,amssymb,amsfonts,amsthm}
\usepackage{amsbsy} 
\usepackage{epsfig}
\usepackage{latexsym}
\input amssym.def
\input amssym.tex

\usepackage{hyperref}

\newcommand{\oarX}[1]{\href{http://arxiv.org/abs/#1}{{\ttfamily #1}}}
\newcommand{\arX}[1]{\href{http://arxiv.org/abs/#1}{{\ttfamily arXiv:#1}}}

\def\barr{\begin{array}}
\def\earr{\end{array}}
\def\half{\frac{1}{2}}
\def\ben{\begin{equation}}
\def\een{\end{equation}}
\def\bs{\begin{subequations}}
\def\es{\end{subequations}}
\def\bena{\begin{eqnarray}}
\def\eena{\end{eqnarray}}

\def\SU{{\rm SU}}

\def\im{{\rm i}}

\def\be{\begin{equation}}
\def\ee{\end{equation}}

\def\bes{\begin{eqnarray}}
\def\ees{\end{eqnarray}}

\begin{document}

\title{Addendum to ``Relational Hamiltonian for group field theory''}

\author{Steffen Gielen}
\affiliation{School of Mathematical Sciences, University of Nottingham, University Park, Nottingham NG7 2RD, United Kingdom}
\email{steffen.gielen@nottingham.ac.uk}\email{axel.polaczek@nottingham.ac.uk}
\author{Axel Polaczek}
\affiliation{School of Mathematical Sciences, University of Nottingham, University Park, Nottingham NG7 2RD, United Kingdom}
\author{Edward Wilson-Ewing}
\affiliation{Department of Mathematics and Statistics, University of New Brunswick,Fredericton, New Brunswick, Canada E3B 5A3}
\email{edward.wilson-ewing@unb.ca}



\begin{abstract}
In this addendum to [Phys.\ Rev.\ D {\bf 99} (2019) 086017], we extend the construction of a Hamiltonian formalism to a new class of group field theory actions with a kinetic term that is local in the group variables and depends only on their derivatives; such a kinetic term will couple magnetic indices of opposite sign in the Peter--Weyl decomposition.  The main results of [Phys.\ Rev.\ D {\bf 99} (2019) 086017] for the resulting cosmology extend to this case.
\end{abstract}


\maketitle

Group field theories (GFTs) \cite{GFT1, GFT2, GFT3} are a non-perturbative approach to quantum gravity whose dynamics are usually given in Lagrangian form.  GFTs are closely related to the spin foam approach to quantum gravity: in the perturbative expansion of a GFT path integral into Feynman diagrams, each Feynman graph is interpreted as a spin foam or discrete quantum gravity amplitude \cite{GFTspinf}. A GFT is defined on an abstract group manifold rather than spacetime, and so the connection to canonical approaches to quantum gravity whose Hamiltonian and momentum constraints encode covariance under space-time diffeomorphisms is indirect \cite{GFT2ndq}.

In canonical quantum gravity, including appropriate matter fields allows for deparametrization which can be much simpler than working in a constrained formalism.  By using matter fields as relational coordinates, it is often possible to obtain a Hamiltonian generating ``true'' time evolution with respect to a matter clock, a famous example is the Brown--Kucha\v{r} dust model \cite{brownkuch}. The idea of deparametrization is also familiar in quantum cosmology, where a common choice of matter clock is a free massless scalar field whose classical evolution is monotonic. 

GFT models for quantum gravity coupled to a massless scalar field have been successfully applied to cosmology \cite{GFTcosmo1, GFTcosmo2}.  In this context, it is possible to define a relational 3-volume observable $V(\chi)$ representing the spatial volume in a GFT state at a given value of the massless scalar ``clock'' $\chi$, similar to what is done in conventional quantum cosmology, e.g., in loop quantum cosmology. Equations for the expectation value of $V(\chi)$ and its derivatives then characterize the dynamics of a homogeneous, isotropic universe and define effective Friedmann equations from GFT \cite{GFTreview}.

Using a massless scalar field $\chi$ as a relational clock in GFT is possible also at a more general level:
interpreting derivatives of a GFT field with respect to $\chi$ as velocities, a Hamiltonian formalism can be constructed for a class of GFT actions, while canonical commutation relations for the field and momentum operators follow directly from the structure of the GFT action \cite{hamiltGFT} (rather than being postulated as in previous work).

In this addendum we extend the construction of a Hamiltonian formalism for GFT to a new class of actions that, when written in the Peter--Weyl decomposition, couple field modes with opposite magnetic indices.  This occurs in GFT actions if the kinetic term is local in the group variables $g_a$ and only depends on derivatives of these group variables (and potentially depends on the scalar field $\chi$ as well); this is closely analogous to what occurs for a free real scalar field in $\mathbb{R}^{d,1}$ whose Lagrangian is local in position space, but takes the form
\ben
\mathcal{L}[\phi]=\frac{1}{2}\dot\phi(\vec{k})\dot\phi(-\vec{k})-\frac{|\vec{k}|^2}{2}\phi(\vec{k})\phi(-\vec{k})
\een
in momentum space. 

In the following we consider theories with a real GFT field $\varphi$ corresponding to quantum gravity with gauge group $\SU(2)$ minimally coupled to a massless scalar field $\chi$ and with an action of the general form
\ben
S[\varphi]=K[\varphi]-V[\varphi],
\een
where $K[\varphi]$ is the kinetic term quadratic in $\varphi$ and $V[\varphi]$ contains higher order terms. $\varphi$ is a real-valued function on $\SU(2)^4\times\mathbb{R}$ with an additional ``gauge invariance''
\ben
\varphi(g_1,g_2,g_3,g_4,\chi)=\varphi(g_1h,g_2h,g_3h,g_4h,\chi)
\een
for any $h\in\SU(2)$. Using the shorthand $\varphi(\vec{g},\chi)\equiv\varphi(g_1,g_2,g_3,g_4,\chi)$, the Peter--Weyl decomposition of such a $\varphi$ into modes labeled by $\SU(2)$ representation data is
\ben
\!\! \varphi(\vec{g},\chi)= \!\!\! \sum_{j_i,m_i,n_i,\iota} \!\!\!\! \varphi^{\vec{\jmath},\iota}_{\vec{m}} \!(\chi)\,\mathcal{I}^{\vec{\jmath},\iota}_{\vec{n}}\prod_{a=1}^4 \! \sqrt{2j_a+1}\, D^{j_a}_{m_a n_a}(g_a),
\een
which is Eq.~(2) in Ref.~\cite{hamiltGFT} apart from the different normalization factor in front of the Wigner matrices $D^j_{mn}(g)$. The sum is over $\SU(2)$ irreducible representations $j_i$, intertwiners $\mathcal{I}$ labeled by $\iota$ and magnetic indices $m_i$ and $n_i$ taking values from $-j_i$ to $j_i$.  

Note that while $\varphi(\vec{g},\chi)$ is real-valued, the modes $\varphi^{\vec{\jmath},\iota}_{\vec{m}}(\chi)$ in the Peter--Weyl decomposition are complex-valued. The reality conditions for these modes are obtained from the identity
\ben
\overline{D^j_{mn}(g)}=(-1)^{m-n}D^j_{-m,-n}(g),
\een
as well as the intertwiner relation
\ben
\mathcal{I}^{\vec{\jmath},\iota}_{\vec{n}}=(-1)^{\sum_i j_i} \mathcal{I}^{\vec{\jmath},\iota}_{-\vec{n}},
\een
together with the fact that an intertwiner is only nonvanishing if $\sum_i n_i=0$. The result is \cite{ooguri}
\ben
\overline{\varphi^{\vec{\jmath},\iota}_{\vec{m}}(\chi)}=(-1)^{\sum_i (j_i-m_i)}\varphi^{\vec{\jmath},\iota}_{-\vec{m}}(\chi).
\label{realityc}
\een
(See Ref.~\cite{primer} for a review of $SU(2)$ recoupling theory.)  This is analogous to the reality condition $\overline{\phi(\vec{k})}=\phi(-\vec{k})$ for the Fourier modes of a scalar field in $\mathbb{R}^{d,1}$, although with an extra sign factor.

We now focus on a class of kinetic terms $K[\varphi]$ in the action (as in Ref.~\cite{hamiltGFT}, the specific form of $V[\varphi]$ will not be used in what follows, and so for simplicity we only consider the case of vanishing $V[\varphi]$). The Ooguri model \cite{ooguri} which defines a GFT for topological BF theory has
\ben
K[\varphi]= \half \sum_{\vec{\jmath},\vec{m},\iota} |\varphi^{\vec{\jmath},\iota}_{\vec{m}}|^2
\een
but more general forms, in particular including derivatives with respect to the $g_a$ in the original group representation, are often of interest. We assume a general type of kinetic term of the form \cite{GFTcosmo1, GFTcosmo2}
\ben
K[\varphi]= \int{\rm d}\chi \sum_{\vec{\jmath},\vec{m},\iota} \overline{\varphi^{\vec{\jmath},\iota}_{\vec{m}}(\chi)} \, \mathcal{K}_{\vec{\jmath},\vec{m},\iota}(\chi) \, \varphi^{\vec{\jmath},\iota}_{\vec{m}}(\chi),
\label{kinetic}
\een
with $\mathcal{K}_{\vec{\jmath},\vec{m},\iota}(\chi)$ real-valued to ensure the action is real.

This form of $K[\varphi]$ is motivated by actions whose kinetic term is local in the group variables, and depends only on their derivatives.  For example, for the kinetic term \cite{lowspin}
\ben
K[\varphi] = \! \int \! {\rm d}\chi \, {\rm d}^4 g \:\, \varphi(\vec{g}, \chi) \left( \mu + \alpha \sum_a \Delta_{g_a} + \beta \partial_\chi^2 \right) \varphi(\vec{g}, \chi), \nonumber
\een
with $\Delta_{g_a}$ the Laplace--Beltrami derivative operator acting on the $a^{\rm th}$ $\SU(2)$ argument of $\varphi$, then $\mathcal{K}_{\vec{\jmath},\vec{m},\iota}(\chi) = \mu - \alpha \sum_a j_a(j_a+1) + \beta \partial_\chi^2$.  To avoid any loss of generality, in the following we consider the kinetic term \eqref{kinetic}.

The $\mathcal{K}_{\vec{\jmath},\vec{m},\iota}(\chi)$ term in Eq.~\eqref{kinetic} can be written as a (potentially infinite) expansion in derivatives with respect to $\chi$ \cite{GFTcosmo1},
\ben
\mathcal{K}_{\vec{\jmath},\vec{m},\iota}(\chi) = \mathcal{K}_{\vec{\jmath},\vec{m},\iota}^{(0)} + \mathcal{K}_{\vec{\jmath},\vec{m},\iota}^{(2)}\partial_\chi^2 + \ldots
\label{expansion}
\een
We then truncate Eq.~(\ref{expansion}) after the second term as in previous work \cite{GFTcosmo1, GFTcosmo2, hamiltGFT} as higher order terms will be suppressed by the Planck scale.  With this truncation, and absorbing some numerical factors, the kinetic term becomes
\ben
K[\varphi]= \half\int{\rm d}\chi \sum_{\vec{\jmath},\vec{m},\iota} \varphi^{\vec{\jmath},\iota}_{-\vec{m}}(\chi)\left[\mathcal{K}_{\vec{\jmath},\vec{m},\iota}^{(0)}+\mathcal{K}_{\vec{\jmath},\vec{m},\iota}^{(2)}\partial_\chi^2\right]\varphi^{\vec{\jmath},\iota}_{\vec{m}}(\chi)\,.
\label{kinetictr}
\een
Notice that without loss of generality we can take $\mathcal{K}_{\vec{\jmath}, \vec{m}, \iota} = \mathcal{K}_{\vec{\jmath}, -\vec{m}, \iota}$ in the following. After a Legendre transform as defined in Ref.~\cite{hamiltGFT}, the GFT Hamiltonian is 
\ben
\mathcal{H}=-\half\sum_{\vec{\jmath},\vec{m},\iota}\left[\frac{\pi^{\vec{\jmath},\iota}_{\vec{m}}\pi^{\vec{\jmath},\iota}_{-\vec{m}}}{\mathcal{K}_{\vec{\jmath},\vec{m},\iota}^{(2)}}+\mathcal{K}_{\vec{\jmath},\vec{m},\iota}^{(0)}\varphi^{\vec{\jmath},\iota}_{\vec{m}}\varphi^{\vec{\jmath},\iota}_{-\vec{m}}\right]
\een
where the elementary Poisson brackets at equal time are
\ben
\{\varphi^{\vec{\jmath},\iota}_{\vec{m}}(\chi),\pi^{\vec{\jmath}',\iota'}_{\vec{m}'}(\chi)\}=\delta^{\vec{\jmath},\vec{\jmath}'}\delta_{\vec{m},\vec{m}'}\delta^{\iota,\iota'}\,.
\een
One can replace $\varphi$ and $\pi$ by a complex variable $a_{\vec{\jmath},\vec{m},\iota}$ defined by
\bena
\!\!\!\!\!\!\!\!\! a_{\vec{\jmath},\vec{m},\iota} & = & \frac{1}{\sqrt{2\hbar \omega^{\vec{\jmath},\iota}_{\vec{m}}}} \left(\omega^{\vec{\jmath},\iota}_{\vec{m}}\varphi^{\vec{\jmath},\iota}_{\vec{m}}+\im (-1)^{\sum_i (j_i-m_i)}\pi^{\vec{\jmath},\iota}_{-\vec{m}} \right) \!,
\\
\!\!\!\!\!\!\!\!\! \bar{a}_{\vec{\jmath},\vec{m},\iota} & = & \frac{1}{\sqrt{2\hbar \omega^{\vec{\jmath},\iota}_{\vec{m}}}} \left((-1)^{\sum_i (j_i-m_i)}\omega^{\vec{\jmath},\iota}_{\vec{m}}\varphi^{\vec{\jmath},\iota}_{-\vec{m}}-\im \pi^{\vec{\jmath},\iota}_{\vec{m}} \right) \!,
\eena
where
\ben
\omega^{\vec{\jmath},\iota}_{\vec{m}}\equiv\sqrt{\left|\mathcal{K}_{\vec{\jmath},\vec{m},\iota}^{(0)}\mathcal{K}_{\vec{\jmath},\vec{m},\iota}^{(2)}\right|}\,.
\een
$a_{\vec{\jmath},\vec{m},\iota}$ and $\bar{a}_{\vec{\jmath},\vec{m},\iota}$ are complex conjugates and satisfy
\ben
\{a_{\vec{\jmath},\vec{m},\iota}(\chi),\bar{a}_{\vec{\jmath}',\vec{m}',\iota'}(\chi)\}=\frac{1}{\im\hbar}\delta_{\vec{\jmath},\vec{\jmath}'}\delta_{\vec{m},\vec{m}'}\delta_{\iota,\iota'}\,.
\een
After canonical quantization, the corresponding creation and annihilation operators then satisfy 
\ben
[\hat{a}_{\vec{\jmath},\vec{m},\iota},\hat{a}^\dagger_{\vec{\jmath}',\vec{m}',\iota'}]=\delta_{\vec{\jmath},\vec{\jmath}'}\delta_{\vec{m},\vec{m}'}\delta_{\iota,\iota'}\,.
\een
Writing the quantum Hamiltonian as
\ben
\hat{\mathcal{H}} = \sum_{\vec{\jmath},\vec{m},\iota} \hat{\mathcal{H}}_{\vec{\jmath},\vec{m},\iota}\,,
\een
the contributions $\hat{\mathcal{H}}_{\vec{\jmath},\vec{m},\iota}$ for each $(\vec{\jmath},\vec{m},\iota)$ take different forms depending on the signs of $\mathcal{K}_{\vec{\jmath},\vec{m},\iota}^{(0)}$ and $\mathcal{K}_{\vec{\jmath},\vec{m},\iota}^{(2)}$. If these coefficients have the same sign for given $(\vec{\jmath},\vec{m},\iota)$,
\ben
\!\!\!\! \hat{\mathcal{H}}_{\vec{\jmath},\vec{m},\iota} = (-1)^{\sum_i (j_i-m_i)}\hbar M_{\vec{\jmath},\vec{m},\iota}\left(\hat{a}^\dagger_{\vec{\jmath},\vec{m},\iota}\hat{a}_{\vec{\jmath},\vec{m},\iota}+\half\right)
\een
with
\ben
M_{\vec{\jmath},\vec{m},\iota} = - {\rm sgn}(\mathcal{K}_{\vec{\jmath},\vec{m},\iota}^{(0)}) \sqrt{\left| \frac{\mathcal{K}_{\vec{\jmath},\vec{m},\iota}^{(0)}}{\mathcal{K}_{\vec{\jmath},\vec{m},\iota}^{(2)}} \right|\, }\,,
\een
whereas if their signs differ,
\ben
\hat{\mathcal{H}}_{\vec{\jmath},\vec{m},\iota} = \frac{\hbar}{2}M_{\vec{\jmath},\vec{m},\iota}\left(\hat{a}^\dagger_{\vec{\jmath},\vec{m},\iota}\hat{a}^\dagger_{\vec{\jmath},-\vec{m},\iota}+\hat{a}_{\vec{\jmath},\vec{m},\iota}\hat{a}_{\vec{\jmath},-\vec{m},\iota}\right)\,.
\label{squeezingh}
\een
As discussed in Ref.~\cite{hamiltGFT}, the Hamiltonian (\ref{squeezingh}) does not leave the Fock vacuum invariant, but will generate excitations; time evolution for Eq.~(\ref{squeezingh}) corresponds to a squeezing of the Fock vacuum (see also Ref.~\cite{toym}). Excitations are created pairwise, with opposite values for the magnetic indices, which is the analog of creating particles with opposite momenta in usual scalar field theory. This coupling of modes with opposite magnetic indices is the only new feature in our analysis here when compared to the results of Ref.~\cite{hamiltGFT}. 

In the cosmological interpretation of GFT dynamics, squeezing corresponds to exponential expansion compatible with the classical Friedmann equations and therefore we will focus on Hamiltonians of the form (\ref{squeezingh}), for which the evolution equations are
\bena
\frac{{\rm d}}{{\rm d}\chi}\langle\hat{a}_{\vec{\jmath},\vec{m},\iota}\rangle &=& -\im M_{\vec{\jmath},\vec{m},\iota}\langle \hat{a}^\dagger_{\vec{\jmath},-\vec{m},\iota}\rangle\,,
\\\frac{{\rm d}}{{\rm d}\chi}\langle\hat{a}^\dagger_{\vec{\jmath},\vec{m},\iota}\rangle &=& \im M_{\vec{\jmath},\vec{m},\iota}\langle \hat{a}_{\vec{\jmath},-\vec{m},\iota}\rangle\,.
\eena
The general solution to these equations is 
\bena
\langle \hat{a}_{\vec{\jmath},\vec{m},\iota} \rangle & = & a^0_{\vec{\jmath},\vec{m},\iota} \cosh(M_{\vec{\jmath},\vec{m},\iota}\chi) - \im \bar{a}^0_{\vec{\jmath},-\vec{m},\iota} \sinh(M_{\vec{\jmath},\vec{m},\iota}\chi)\,,\nonumber
\\\langle \hat{a}^\dagger_{\vec{\jmath},\vec{m},\iota} \rangle & = & \bar{a}^0_{\vec{\jmath},\vec{m},\iota} \cosh(M_{\vec{\jmath},\vec{m},\iota}\chi) + \im a^0_{\vec{\jmath},-\vec{m},\iota} \sinh(M_{\vec{\jmath},\vec{m},\iota}\chi)\,.\nonumber
\eena

In a mean-field approximation (expected to be valid for a condensate corresponding to the cosmological sector of a GFT for quantum gravity \cite{GFTreview}), the occupation number per mode is given by
\ben
\langle \hat{a}^\dagger_{\vec{\jmath},\vec{m},\iota}\hat{a}_{\vec{\jmath},\vec{m},\iota} \rangle \approx |\langle \hat{a}_{\vec{\jmath},\vec{m},\iota} \rangle|^2\,.
\een
The symmetric combination $S_{\vec{\jmath},\vec{m},\iota} = \langle \hat{a}^\dagger_{\vec{\jmath},\vec{m},\iota}\hat{a}_{\vec{\jmath},\vec{m},\iota} \rangle+\langle \hat{a}^\dagger_{\vec{\jmath},-\vec{m},\iota}\hat{a}_{\vec{\jmath},-\vec{m},\iota} \rangle$ contains all of the information needed for cosmological applications since the GFT operator corresponding to the spatial volume is insensitive to the values of the magnetic indices. For this quantity we find
\bena
S_{\vec{\jmath},\vec{m},\iota} &=& |\langle \hat{a}_{\vec{\jmath},\vec{m},\iota} \rangle|^2 + |\langle \hat{a}_{\vec{\jmath},-\vec{m},\iota} \rangle|^2 \nonumber \\
&=&\left(|a^0_{\vec{\jmath},\vec{m},\iota}|^2+|a^0_{\vec{\jmath},-\vec{m},\iota}|^2\right) \cosh(2M_{\vec{\jmath},\vec{m},\iota}\chi)\nonumber
\\&& - 2\,{\rm Im}(a^0_{\vec{\jmath},\vec{m},\iota}a^0_{\vec{\jmath},-\vec{m},\iota}) \sinh(2M_{\vec{\jmath},\vec{m},\iota}\chi)\nonumber
\\ &=& A_{\vec{\jmath},\vec{m},\iota}\cosh\big(2M_{\vec{\jmath},\vec{m},\iota}(\chi-\tilde\chi^0_{\vec{\jmath},\vec{m},\iota})\big)
\label{result}
\eena
for suitable constants $A_{\vec{\jmath},\vec{m},\iota}\ge 0$ and $\tilde\chi^0_{\vec{\jmath},\vec{m},\iota}$. Note that the last equality follows from the fact that, after the second equality, the prefactor to the $\cosh(2M_{\vec{\jmath},\vec{m},\iota}\chi)$ term is positive and larger in amplitude (as follows from $|a^0_{\vec{\jmath},\vec{m},\iota} \pm \im \, \bar a^0_{\vec{\jmath},-\vec{m},\iota}|^2 \ge 0$) than the prefactor to the $\sinh(2M_{\vec{\jmath},\vec{m},\iota}\chi)$ term (which can be negative).

Apart from the fact that we are summing the occupation numbers for $(\vec{\jmath},\vec{m},\iota)$ and $(\vec{\jmath},-\vec{m},\iota)$, Eq.~(\ref{result}) reproduces the result of Eq.~(46) in Ref.~\cite{hamiltGFT}. The spatial volume for each mode is given by the occupation number times a volume eigenvalue; the results of Ref.~\cite{hamiltGFT} for effective cosmology, recovery of the classical Friedmann equations at late times and singularity resolution by a bounce then extend to the class of kinetic terms (\ref{kinetictr}). There is a difference in the freedom to choose initial conditions in that the integration constants $A_{\vec{\jmath},\vec{m},\iota}$ and $\tilde\chi^0_{\vec{\jmath},\vec{m},\iota}$ depend on initial conditions for both $(\vec{\jmath},\vec{m},\iota)$ and $(\vec{\jmath},-\vec{m},\iota)$. One might prefer to use states that reflect the symmetry $(\vec{\jmath},\vec{m},\iota)\leftrightarrow(\vec{\jmath},-\vec{m},\iota)$ of the theory, for which $a^0_{\vec{\jmath},\vec{m},\iota}=a^0_{\vec{\jmath},-\vec{m},\iota}$ and then $\langle \hat{a}_{\vec{\jmath},\vec{m},\iota} \rangle = \langle \hat{a}_{\vec{\jmath},-\vec{m},\iota} \rangle$ at all times.

\bigskip

{\em Acknowledgments.} --- The work of S.G.\ was funded by the Royal Society under a Royal Society University Research Fellowship (UF160622) and a Research Grant for Research Fellows (RGF\textbackslash R1\textbackslash 180030), and E.W.-E.\ is supported by funding from the Natural Science and Engineering Research Council of Canada.

\raggedright


\begin{thebibliography}{99}

\bibitem{GFT1} D.~Oriti, The group field theory approach to Quantum Gravity, in {\em Approaches to Quantum Gravity -- Toward a New Understanding of Space, Time and Matter}, edited by D.~Oriti (Cambridge: Cambridge University Press, 2009), \oarX{gr-qc/0607032}.

\bibitem{GFT2} L.~Freidel, Group Field Theory: An Overview, {\em Int.\ J.\ Theor.\ Phys.}  {\bf 44} (2005) 1769--1783, \oarX{hep-th/0505016}.

\bibitem{GFT3} T.~Krajewski, Group field theories {\em PoS QGQGS} {\bf 2011} (2011) 005, \arX{1210.6257}

\bibitem{GFTspinf} M.~P.~Reisenberger and C.~Rovelli, Spacetime as a Feynman diagram: the connection formulation, {\em Class.\ Quant.\ Grav.}  {\bf 18} (2001) 121--140, \oarX{gr-qc/0002095}.

\bibitem{GFT2ndq} D.~Oriti, Group field theory as the second quantization of loop quantum gravity, {\em Class.\ Quant.\ Grav.}  {\bf 33} (2016)  085005, \arX{1310.7786}.

\bibitem{brownkuch} J.~D.~Brown and K.~V.~Kucha\v{r}, Dust as a standard of space and time in canonical quantum gravity, {\em Phys.\ Rev.\ D} {\bf 51} (1995) 5600--5629, \oarX{gr-qc/9409001}.

\bibitem{GFTcosmo1} D.~Oriti, L.~Sindoni and E.~Wilson-Ewing, Emergent Friedmann dynamics with a quantum bounce from quantum gravity condensates, {\em Class.\ Quant.\ Grav.}  {\bf 33} (2016) 224001, \arX{1602.05881}.

\bibitem{GFTcosmo2} D.~Oriti, L.~Sindoni and E.~Wilson-Ewing, Bouncing cosmologies from quantum gravity condensates, {\em Class.\ Quant.\ Grav.} {\bf 34} (2017)  04LT01, \arX{1602.08271}.
 
\bibitem{GFTreview} S.~Gielen and L.~Sindoni, Quantum Cosmology from Group Field Theory Condensates: a Review, {\em SIGMA} {\bf 12} (2016) 082, \arX{1602.08104}.

\bibitem{hamiltGFT} E.~Wilson-Ewing, Relational Hamiltonian for group field theory, {\em Phys.\ Rev.\ D} {\bf 99} (2019) 086017, \arX{1810.01259}

\bibitem{ooguri} H.~Ooguri, Topological lattice models in four dimensions, {\em Mod.\ Phys.\ Lett.\ A} {\bf 7} (1992) 2799--2810, \oarX{hep-th/9205090}.

\bibitem{primer} P.~Martin-Dussaud, A Primer of Group Theory for Loop Quantum Gravity and Spin-foams, \arX{1902.08439}.

\bibitem{lowspin} S.~Gielen, Emergence of a low spin phase in group field theory condensates, {\em Class.\ Quant.\ Grav.}  {\bf 33} (2016) 224002, \arX{1604.06023}.

\bibitem{toym} E.~Adjei, S.~Gielen and W.~Wieland, Cosmological evolution as squeezing: a toy model for group field cosmology, {\em Class.\ Quant.\ Grav.}  {\bf 35} (2018)  105016, \arX{1712.07266}



\end{thebibliography}
\end{document}